\documentclass[aps,prd,natbib,preprint,eqsecnum,nofootinbib,groupedaddress,floatfix,13pt,longbibliography]{revtex4-1}

\usepackage[usenames, dvipsnames]{color}
\usepackage{graphicx,amssymb,amsbsy,amsfonts,amssymb,amsmath}
\usepackage{hyperref}
\usepackage{multirow}
\usepackage{tikz}
\usepackage{stackrel}

\newcommand{\be}{\begin{equation}}
\newcommand{\ee}{\end{equation}}

\newcommand{\bea}{\begin{eqnarray}}
\newcommand{\eea}{\end{eqnarray}}

\newcommand{\eps}{\epsilon}

\def\spa#1.#2{\left\langle#1\,#2\right\rangle}
\def\spb#1.#2{\left[#1\,#2\right]}
\def\spash#1.#2{\spa{\smash{#1}}.{\smash{#2}}}
\def\spbsh#1.#2{\spb{\smash{#1}}.{\smash{#2}}}
\def\sand#1.#2.#3{%
\left\langle\smash{#1}{\vphantom1}^{-}\right|{#2}%
\left|\smash{#3}{\vphantom1}^{-}\right\rangle}
\def\sandpp#1.#2.#3{%
\left\langle\smash{#1}{\vphantom1}^{+}\right|{#2}%
\left|\smash{#3}{\vphantom1}^{+}\right\rangle}
\def\sandpm#1.#2.#3{%
\left\langle\smash{#1}{\vphantom1}^{+}\right|{#2}%
\left|\smash{#3}{\vphantom1}^{-}\right\rangle}
\def\sandmp#1.#2.#3{%
\left\langle\smash{#1}{\vphantom1}^{-}\right|{#2}%
\left|\smash{#3}{\vphantom1}^{+}\right\rangle}

\def\BlackHat{{\sc BlackHat}}
\def\Mathematica{{\sc Mathematica}}

\newbox\charbox
\newbox\slabox
\def\s#1{{      
        \setbox\charbox=\hbox{$#1$}
        \setbox\slabox=\hbox{$/$}
        \dimen\charbox=\ht\slabox
        \advance\dimen\charbox by -\dp\slabox
        \advance\dimen\charbox by -\ht\charbox
        \advance\dimen\charbox by \dp\charbox
        \divide\dimen\charbox by 2
        \raise-\dimen\charbox\hbox to \wd\charbox{\hss/\hss}
        \llap{$#1$} }}

\makeatletter
\def\l@subsubsection#1#2{}
\makeatother
\usepackage{hyperref}

\begin{document}

\hbox{\rm\small
FR-PHENO-2017-025$\null\hskip 9.4cm \null$
UCLA-17-TEP-108
\break}

\title{Planar Two-Loop Five-Gluon Amplitudes from Numerical Unitarity}

\author{
S.~Abreu${}^{a}$, 
F.~Febres Cordero${}^{a}$, 
H.~Ita${}^{a}$, 
B.~Page${}^{a}$ and 
M.~Zeng${}^{b}$
\\
$\null$
\\
${}^a$ Physikalisches Institut, Albert-Ludwigs-Universit\"at Freiburg\\
       D--79104 Freiburg, Germany \\
${}^b$ Mani L.~Bhaumik Institute for Theoretical Physics \\
UCLA Department of Physics and Astronomy \\
Los Angeles, CA 90095, USA
}

\begin{abstract}
We present a calculation of the planar two-loop five-gluon amplitudes.
The amplitudes are obtained in a variant of the generalized unitarity
approach suitable for numerical computations, which we extend for use with finite field arithmetics.
Employing a new method for the generation of unitarity-compatible
integration-by-parts identities, all helicity amplitudes are reduced to a
linear combination of master integrals for the first time. The
approach allows us to compute exact values for the integral
coefficients at rational phase-space points.
All required master integrals are known analytically, and we obtain
arbitrary-precision values for the amplitudes.
\end{abstract}

\maketitle

\tableofcontents
\numberwithin{equation}{section}
\newpage


\section{Introduction}
\label{Sec:Introduction}

The expected precision of upcoming cross-section measurements at the Large Hadron
Collider at CERN currently drives the development of new computational methods in
perturbative quantum-field theory, and in particular in QCD.
At the same time, formal advances in the field of amplitudes are finding
more and more traction and are helping to devise more efficient
approaches to multi-loop computations. 
Here, we aim to contribute to these developments by refining and extending methods to
compute phenomenologically relevant two-loop scattering amplitudes. 
With our results, we advance the state of the art in the calculation of loop
amplitudes with many external particles.
Besides their importance for phenomenology, the new results highlight
the potential of the methods we employ.

In this paper, we present the computation of two-loop gluonic amplitudes which
contribute to three-jet production at hadron colliders at
next-to-next-to-leading
order (NNLO).
Predictions for these processes can be used for constraining
the strong coupling, as it can be extracted
from precision measurements of three- to two-jet-production ratios 
(see e.g.~ref.~\cite{Andersen:2014efa,Badger:2016bpw}).
While the finite parts of our results can be integrated over phase space, significant
additional developments will be required to obtain the full NNLO predictions.
Probably the most challenging step will be the
consistent combination of all virtual and real-radiation contributions
to NNLO accuracy.
Nevertheless, we are optimistic that by providing key ingredients towards this goal, such as the
two-loop scattering amplitudes, we will spur progress in the field to
obtain NNLO predictions for three-jet production in the coming years.

Currently, only a limited set of amplitudes with five or more external
particles are known to two-loop order in QCD. First benchmark results for
two-loop five-gluon amplitudes were obtained for the all-plus-helicity
configuration~\cite{Badger:2013gxa,Badger:2015lda}, whose computation relied on
compact analytic expressions for the integrands. These were computed with
the unitarity method~\cite{Bern:1994zx,Bern:1994cg,Bern:1997sc,Britto:2004nc},
coupled with efficient integrand-reduction techniques
\cite{Ossola:2006us,Mastrolia:2011pr,Mastrolia:2012an,Mastrolia:2012an,Badger:2012dp,Zhang:2012ce}.
The final amplitudes were first obtained by integrating the optimized
integrands numerically and later by explicit reduction to master
integrals~\cite{Gehrmann:2015bfy}. 
The relevant master integrals have been independently evaluated in
ref.~\cite{Papadopoulos:2015jft}.
More recently, the all-plus two-loop amplitudes were obtained from a recursive
approach~\cite{Dunbar:2016aux} that was also extended to six and seven external
gluons~\cite{Dunbar:2016gjb,Dunbar:2017nfy}.
During the final stages of this work, benchmark results for planar
five-gluon amplitudes with generic helicity assignments
were presented~\cite{Badger:2017jhb}.
The amplitudes were integrated with a combination of numeric and analytic
techniques, after partial reduction using unitarity cuts.

In this paper, we apply the numerical unitarity approach \cite{Ossola:2006us,Ellis:2007br,Giele:2008ve,Berger:2008sj}, which was recently extended to two-loops \cite{Ita:2015tya,Abreu:2017idw,Abreu:2017xsl}, to compute the full set of
planar five-gluon two-loop amplitudes.
This computation marks the
first time that a full set of two-loop five-scale amplitudes is
reduced to a linear combination of integral coefficients multiplied by
corresponding master integrals. The set of master integrals is minimal, as all possible integral relations have been imposed.
The potential of this approach has recently been demonstrated
in the context of planar four-gluon two-loop
amplitudes~\cite{Abreu:2017xsl}, and in order to handle the complexity of the five-point
amplitudes we refine it in the present work in two ways.

Firstly, we use an improved approach to obtain analytic expressions for 
integration-by-parts (IBP)
identities which are important in our construction of the amplitude.  The unitarity
method~\cite{Bern:1994zx,Bern:1994cg,Bern:1994cg,Britto:2004nc} matches unitarity cuts to
an ansatz of the loop amplitudes written in terms of master integrals.  When considered
at the integrand level, one-loop unitarity approaches~\cite{Giele:2008ve,
Ossola:2006us} relate cuts to an ansatz for the amplitude's integrand written in terms of surface
terms (spurious numerators that integrate to zero) and master integrands
(the integrands corresponding to the master integrals).
The numerical unitarity method at the two-loop
level~\cite{Ita:2015tya,Abreu:2017xsl, Abreu:2017idw} is a natural
generalization of the one-loop
integrand level approach: the unitarity cuts are matched to a complete parameterization
of the integrand in terms of 
specially tailored IBP relations~\cite{Chetyrkin:1981qh, Laporta:2001dd,
Laporta:1996mq} (the surface terms) and master integrands.
As such, the method directly achieves a reduction to master integrals at the integrand level.
Compared to more canonical approaches, the challenging inversion of the IBP systems required to
obtain reduction tables for tensor integrals~\cite{Laporta:2001dd} is avoided.
While we leave a more detailed discussion of the way we generate the required IBP relations to
future work~\cite{GeoIBP}, we describe the central formal improvements for their
derivation.

Secondly, we adapt numerical unitarity to exact numerical
computations based on the finite field 
approach~\cite{vonManteuffel:2014ixa}
and its application to unitarity cuts~\cite{Peraro:2016wsq}.
The finite field arithmetics allow us to compute exact values for the integral
coefficients at rational phase-space points. 
Combined with the analytic expressions for two-loop master integrals with five massless external
legs~\cite{Papadopoulos:2015jft}, we obtain stable results whose precision only
depends on the number of digits to which the multiple polylogarithms appearing
in the master integrals are evaluated.
The extension to finite fields will be
crucial for a future use of functional reconstruction techniques to
determine analytic forms of the amplitudes from the exact numerical results.

The article is organized as follows. In Section~\ref{sec:Methods} we
summarize the numerical unitarity method, as well as the two
refinements already highlighted. In Section~\ref{sec:Implementation} we give explicit 
details of our implementation.
In Section~\ref{sec:Results} we present our results for the four independent
leading-color helicity amplitudes required for all five-gluon scattering
processes at a given phase-space point.
In Section~\ref{sec:Conclusion} we give our conclusions and
outlook.
Finally, in Appendix~\ref{sec:IR} we discuss the universal infrared structure of the
amplitudes which we use to validate our results.


\section{Numerical Unitarity Approach}
\label{sec:Methods}

The main goal of this paper is the calculation of
two-loop corrections to five-gluon scattering amplitudes
in the leading-color approximation for any helicity configuration in pure Yang-Mills theory.
We write the perturbative expansion
of a bare five-gluon helicity amplitude as
\begin{align}\begin{split}\label{eq:bareAmp}
  \mathcal{A}(\{p_i,h_i\}_{i=1,\ldots,5})\big\vert_{\textrm{leading color}}
  =&
\sum_{\sigma\in S_5/Z_5} {\rm Tr}\left(
    T^{a_{\sigma(1)}} T^{a_{\sigma(2)}}
    T^{a_{\sigma(3)}} T^{a_{\sigma(4)}} T^{a_{\sigma(5)}} \right)\\
    &g^{3}_0
    \left(\mathcal{A}^{(0)}
  +\frac{\alpha_0N_C}{4\pi}\mathcal{A}^{(1)}
  +\left(\frac{\alpha_0N_C}{4\pi}\right)^2\mathcal{A}^{(2)}
  +\mathcal{O}(\alpha_0^3)
  \right)\,,
\end{split}\end{align}
where $\alpha_0=g_0^2/(4\pi)$ is the bare QCD coupling. The set $\sigma$ denotes
all non-cyclic permutations of five indices,
and we make it explicit that the color
structures beyond tree-level give
only a factor of $N_C$ at each order in the expansion
in the leading-color approximation.
The planar amplitudes $\mathcal{A}^{(j)}$ are
functions of the momenta $p_{\sigma(i)}$, helicities $h_{\sigma(i)}$ and
color indices $a_{\sigma(i)}$. In the leading-color approximation there is a single color structure and it
is thus sufficient to specify an helicity assignment for the ordered set of legs,
\begin{align}\begin{split}\label{eq:orderedAmp}
    \mathcal{A}^{(i)}(1^{h_1}, 2^{h_2}, 3^{h_3}, 4^{h_4}, 5^{h_5}) \,.
\end{split}\end{align}
In this section we describe our approach to the calculation of the $\mathcal{A}^{(i)}$.
Although in this paper we focus on the leading-color contributions, the approach we use
is completely generic and applicable beyond this approximation.

\subsection{Overview}
\label{sec:NumUnitarity}

We apply a variant of the unitarity method
\cite{Bern:1994zx,Bern:1994cg,Bern:1997sc,Britto:2004nc} suitable for automated numerical
computations of multi-loop
amplitudes~\cite{Ita:2015tya,Abreu:2017xsl,Abreu:2017idw}. 
Our approach starts
from the standard decomposition of the amplitude in terms of master integrals,
\begin{equation}\label{eq:A}
\mathcal{A}^{(k)}=\sum_{\Gamma\in \Delta} \sum_{i\,\in\, M_\Gamma} c_{\Gamma,i} \,
{\mathcal I}_{\Gamma,i}\,,
\end{equation}
where the index $\Gamma$ labels the different propagator structures and $i$ denotes an
additional degeneracy for cases where multiple master integrals with identical
propagator structures appear, as can happen beyond one loop.
The set $\Delta$ denotes the collection of diagrams which specify the possible
propagator structures of the amplitude, and only depends on the kinematics
of the amplitude (i.e. on the momenta and masses of the particles involved).
In figs.~\ref{fig:PropagatorStructures} and \ref{fig_master_int}, we
give an explicit example for a set of propagator structures $\Delta$ and 
the corresponding set of master integrals together with their multiplicity,
in the case of two-loop five-gluon leading-color amplitudes.
The aim of the computation is to determine the coefficient functions $c_{\Gamma,i}$
which contain the process specific information.

Next, we promote the decomposition of the amplitude of
eq.~\eqref{eq:A} to the integrand level.
The integrand, which we denote ${\cal A}(\ell_l)$ with $\ell_l$ representing the loop momenta
$\ell_1,\ldots,\ell_L$ in a $L$-loop calculation,
is decomposed as~\cite{Ita:2015tya}
\begin{equation}\label{eq:AL}
{\cal A}^{(k)}(\ell_l)=\,\sum_{\Gamma \in \Delta}\,\,\sum_{i\,\in\, M_\Gamma\cup
S_\Gamma} \frac{ c_{\Gamma,i} \,m_{\Gamma,i}(\ell_l)}{\prod_{j\in P_\Gamma}
\rho_j}\,, 
\end{equation}
where the $\rho_j$ are the inverse propagators and $P_\Gamma$ the set of propagators
associated with the topology $\Gamma$. 
Compared with (\ref{eq:A}), we have extended the sum on $i$ to also go over so-called
surface terms 
in the set $S_\Gamma$, which vanish upon integration but are required to
parameterize the integrand. 
We work in dimensional regularization with $D$ space-time dimensions,
and thus allow $D$-dependent surface terms.  
All in all, the
decomposition~\eqref{eq:AL} is rather universal, depending only on the set of
diagrams $\Delta$ and the allowed power counting of the theory under
consideration. As in the decomposition of eq.~\eqref{eq:A}, the spectrum-specific
information is carried by the coefficients $c_{\Gamma,i}$.
Obtaining a complete parameterization of the 
integrand of the amplitude is a non-trivial process, and in the next subsection
we explain how this was achieved. For now, it is sufficient to assume that such a
decomposition exists.

The ansatz (\ref{eq:AL}) holds as a function of the loop momenta $\ell_l$.
Provided one can evaluate  ${\cal A}^{(k)}(\ell_l)$ at specific values of $\ell_l$,
it can be used to construct a system of equations for the coefficients
$c_{\Gamma,i}$.
In a generalized unitarity calculation, the coefficients $c_{\Gamma,i}$ are determined
from a set of \emph{cut equations} obtained by setting
$\ell_l$ to specific on-shell configurations.
Indeed, the leading terms in the various on-shell limits of the ansatz (\ref{eq:AL})
factorize, yielding the cut equations
\begin{eqnarray}
\sum_{\rm states}\prod_{i\in T_\Gamma} {\cal A}^{\rm tree}_i(\ell_l^\Gamma) =
\sum_{\substack{\Gamma' \ge \Gamma\,,\\ i\,\in\,M_{\Gamma'}\cup S_{\Gamma'}} }
\frac{ c_{\Gamma',i}\,m_{\Gamma',i}(\ell_l^\Gamma)}{\prod_{j\in
    (P_{\Gamma'}\setminus P_\Gamma) } \rho_j(\ell_l^\Gamma)}\,. \label{eq:CE} 
\end{eqnarray}
Here, the set $T_\Gamma$ labels all tree amplitudes corresponding to the vertices in
the diagram~$\Gamma$, and the state sum runs over all internal states allowed by the theory.
The $\ell^\Gamma_l$ correspond to a configuration of the loop momenta where the propagators
in the set $P_\Gamma$ associated with the topology $\Gamma$ are on-shell.
Although this limit probes all topologies $\Gamma'$ 
for which $P_\Gamma\subseteq P_{\Gamma'}$,
through a top-down approach, i.e. starting from the topologies with the most propagators,
we can make sure that the only unknowns in eq.~\eqref{eq:CE} are the coefficients
$c_{\Gamma,i}$ belonging to $\Gamma$ itself. By sampling over
enough values of $\ell^\Gamma_l$, we can build a system
of equations big enough to determine all the coefficients. We note that beyond one loop
there are topologies that correspond to subleading terms in the on-shell limit $\ell^\Gamma_l$,
in which case no factorization of the ansatz (\ref{eq:AL}) is known. This is a minor
obstacle that can be easily overcome \cite{Abreu:2017idw}.

For the process we are concerned with in this paper, two-loop gluonic
amplitudes in the leading-color approximation, the state sum in
eq.~\eqref{eq:CE} goes over the $(D_s-2)$ gluon helicity
states. Furthermore, as the surface terms are functions of the
space-time dimension parameter $D$, the coefficients $c_{\Gamma,i}$
obtained by solving the cut equations are functions of both $D_s$ and
$D$.
The value chosen for $D_s$ defines different regularization schemes. We have implemented both
the 't Hooft-Veltman (HV) scheme \cite{tHooft:1972tcz} where $D_s=D=4-2\epsilon$
and the four-dimensional helicity
(FDH) scheme \cite{Bern:1991aq,Bern:2002zk} where $D_s=4$.

\subsection{Efficient Construction of Unitarity-Compatible IBP Identities}
\label{sec:IBP}

We now return to the discussion of how the decomposition in
eq.~\eqref{eq:AL} is achieved. Here we present the generic framework
of our method but leave more specific details for a future dedicated
publication \cite{GeoIBP}.
The surface terms are constructed from a complete set of so-called {\it
unitarity-compatible} integration-by-parts (IBP) identities
\cite{Gluza:2010ws,Schabinger:2011dz,Ita:2015tya,Larsen:2015ped}.
These relations are built such that they do not involve any new propagator structures
besides those in $\Delta$, the full set of propagator structures we started with.
As is the case in more canonical approaches to multi-loop amplitude calculations,
our approach requires a complete set of 
IBP identities for each of the
propagator structures~\cite{Abreu:2017xsl}.
In the context of the ansatz in eq.~\eqref{eq:AL},
completeness means that surface terms and master integrands span the full function space prescribed by the power counting of the theory.
The system of equations in eq.~\eqref{eq:CE} is then invertible and by solving it
we achieve a full reduction of the amplitude to master integrals.

In principle, the desired decomposition can be achieved by
IBP reducing all allowed insertions of irreducible scalar products (ISPs)~\footnote{Precisely,
ISPs are scalar products involving loop
momenta that cannot be expressed solely in terms of inverse propagators.}
on a given propagator structure. The reduction identities may then be
used as the set of surface terms in eq.~(\ref{eq:AL}).
In practice, the currently available reduction
programs~\cite{Anastasiou:2004vj,Smirnov:2008iw,Smirnov:2014hma,Studerus:2009ye,vonManteuffel:2012np,Maierhoefer:2017hyi}
struggle with the high tensor-rank appearing in two-loop five-gluon
amplitudes as the size of the IBP system that needs to be solved
becomes prohibitive.
Instead, we follow an alternative path to obtain unitarity-compatible relations
which has recently received much attention~\cite{Gluza:2010ws,
Schabinger:2011dz, Ita:2015tya, Larsen:2015ped, Georgoudis:2016wff,
Ita:2016oar, Zhang:2016kfo, Abreu:2017xsl, Bern:2017gdk}. By only
generating unitarity-compatible IBP relations from the start, we
ensure that the size of the system remains manageable.
In our approach, the solution of the IBP system is done
by solving the cut equations \eqref{eq:CE}, which allows to efficiently
organize the calculation.

We start from the construction of the `transverse IBP identities' from
refs.~\cite{Ita:2015tya,Abreu:2017xsl}, but further improve the construction of
the remaining identities by solving syzygy module equations. These are closely
related to the defining equation for the so-called IBP-generating vectors
$\{u^\nu_k\}$~\cite{Gluza:2010ws},
\begin{equation} \label{eq:GKK}
u_k^\nu \frac{\partial}{\partial \ell_k^\nu} \rho_j = f_j \rho_j, \quad \forall j \in P_\Gamma,
\end{equation}
where $1 \leq k \leq L$ is the loop momentum label, $\nu$ is the Lorentz
index and there is no summation over $j$. We require $f_j$ to be polynomial in the dot
products built from loop momenta and external momenta. The vectors $\{u_k^\nu\}$ are required to be polynomial 
vectors in the loop momenta. The above equation
ensures that the IBP relation
\begin{equation}\label{eq:ibp-generic}
    \int \prod_{l=1,L} d^D\ell_l \sum_k\frac{\partial}{\partial \ell^\nu_k}
    \left[\frac{u^\nu_k}{\prod_{j\in P_\Gamma}\rho_j}\right]=0\,
\end{equation}
contains no integrals with raised propagator powers and is therefore a suitable
unitarity-compatible IBP relation. In the context of eq.~(\ref{eq:AL}),
the left-hand-side of \eqref{eq:ibp-generic} gives a valid surface term.

Our method uses inverse propagator variables to trivialize the separation of terms that vanish on-shell from ISPs that survive on the cut, as in~\cite{Ita:2015tya,Abreu:2017xsl}.
However, we further reduce the polynomial degree of the syzygy equations by simply
rewriting the defining equation for IBP-generating vectors (\ref{eq:GKK}) in terms of the components of the loop and the external momenta.
If starting from a formulation in
terms of so-called  Baikov polynomials \cite{Larsen:2015ped},
this method of solving eq.~\eqref{eq:GKK} may be viewed as a variant of the
intersection method~\cite{Zhang:2016kfo}.
We write down the following ansatz for the
IBP-generating vector
\begin{equation} u_k^\nu \frac{\partial}{\partial \ell_k^\nu}= \left( u^{\rm
loop}_{ka} \ell_a^\nu +  u^{\rm ext}_{kc} p_c^\nu \right)
\frac{\partial}{\partial \ell_k^\nu}, \end{equation}
where we have an implicit sum over the label $a$ for independent loop momenta
and label $c$ for independent external momenta. 
Consistent with the fact that the original vectors $\{ u_k^\nu \}$ are polynomial we require 
the vector components $u^{\rm loop}_{ka}$ and $u^{\rm
ext}_{kc}$ to be a polynomial in the dot products built from loop momenta and
external momenta. 
Since the latter can be
expressed in terms of ISPs and inverse propagators, the variables $u^{\rm
loop}_{ka}$, $u^{\rm ext}_{kc}$ and $f_j$ are polynomials in ISPs and inverse
propagators. 
Eq.\ \eqref{eq:GKK} then becomes
\begin{equation} \label{eq:moduleSyz}
\left( u^{\rm loop}_{ka} \ell_a^\nu +  u^{\rm ext}_{kc} p_c^\nu \right)
\frac{\partial}{\partial \ell_k^\nu} 
\begin{pmatrix} \rho_{j(1)} \\ \rho_{j(2)} \\ \vdots \\ \rho_{j(|\Gamma|)}
\end{pmatrix} - 
\begin{pmatrix} f_{j(1)} \rho_{j(1)} \\ f_{j(2)} \rho_{j(2)} \\ \vdots \\
f_{j(|\Gamma|)} \rho_{j(|\Gamma|)} \end{pmatrix}
=0\,, \end{equation}
where there is implicit summation over $a$, $c$, $k$, and $\nu$, and the inverse
propagator labels $j(i)$ run over all propagators in the set $P_\Gamma$,
\begin{equation} \{ j(1),j(2), \dots, j(|\Gamma|) \} \equiv P_\Gamma \, .
\end{equation}
Note that both
\begin{equation} \ell_a^\nu \frac{\partial}{\partial \ell_k^\nu} \rho_j
\qquad\qquad
\textrm{and}
\qquad\qquad
 p_c^\nu \frac{\partial}{\partial \ell_k^\nu} \rho_j
\end{equation}
evaluate to contractions of the propagator momenta with loop and external 
momenta, and can be expressed as matrices of linear polynomials in ISPs and inverse
propagators.
Similarly the second term in eq.~\eqref{eq:moduleSyz} can be expressed
in matrix form with propagator variables on the diagonal. Consequently
eq.~\eqref{eq:moduleSyz} has the form of a \emph{syzygy equation over
a module} for the unknown polynomials $u^{\rm loop}_{ka}$, $u^{\rm
ext}_{kc}$ and $f_j$. The equations are defined over the freely
generated polynomial ring given by ISPs and inverse propagators.
Syzygy equations can be solved using algorithms in
computational algebraic geometry implemented in e.g.~the {\sc Singular}
computer algebra system \cite{DGPS}. We will give more details of how we solved
the syzygy equations using {\sc Singular} in the next section.

Once a generating set of vector components $u^{\rm loop}_{ka}$ and $u^{\rm
ext}_{kc}$ has been obtained, a sufficient set of IBP relations is obtained by
multiplying the generators with irreducible numerators. 
At this stage, we explicitly impose off-shell power-counting conditions
(i.e.~we consider both on-shell and propagator variables),
sometimes after forming linear combinations of IBP relations.  In a second step
the independence of IBP identities has to be established. Given the manifest
dependence on propagator variables and the vectors compatibility with unitarity
cuts, it is natural to switch between on-shell and off-shell as well as
numerical and analytic perspectives to simplify computational
steps~\cite{Ita:2015tya,Larsen:2015ped}.  In particular, the 
validation of the linear independence of the off-shell relations
\cite{Ita:2015tya,Abreu:2017xsl} is most easily performed by setting
propagator variables to their on-shell values, as redundancy of the
relations is then manifest.
Finally, we note it is often convenient to count the number of master integrals
on-shell~\cite{Ita:2015tya,Larsen:2015ped,Georgoudis:2016wff,Abreu:2017xsl}.

\subsection{Numerical Unitarity over an Arbitrary Number Field}
\label{sec:FiniteFields}

Numerical computations have better scaling properties than analytic
ones for processes with a large number of scales.
However, the
stability and efficiency of analytic results for amplitudes often offer major
advantages over numerical evaluations,
specially when used in conjunction with Monte Carlo programs.
The boundary between the two approaches is becoming blurred in
the field of scattering amplitude calculations with
the recent introduction of functional reconstruction techniques
\cite{vonManteuffel:2014ixa,Peraro:2016wsq,Abreu:2017xsl}. In this approach, numerical
evaluations are used to completely reconstruct analytic expressions
from numerical samples. Whilst already successful in the
reconstruction of two-loop amplitudes with two scales
\cite{Abreu:2017xsl}, it is clear that multi-scale problems such as the
five-gluon amplitudes are
significantly more complicated. Not only must more efficient techniques be
used for the functional reconstruction \cite{Peraro:2016wsq},  but
the loss of precision associated with
floating-point arithmetics in multi-scale computations
poses a considerable difficulty for the reconstruction procedure.
To address this issue, in this
section we reformulate numerical unitarity for gluonic
amplitudes so that it only employs operations defined on a field
(addition, subtraction, multiplication and division).
In this way, we
open the door to the use of exact arithmetics, such as those of
rational numbers or finite fields. This eliminates any question of
numerical stability, whilst still leveraging the speed of modern
computer hardware.

The extension of numerical unitarity to employ only field
operations is non-trivial. Indeed, operations such
as taking square roots or, more generally, solving generic polynomial equations
are not allowed, and these feature heavily
in many formulations of generalized unitarity.
Here, we build on the ideas proposed in ref.~\cite{Peraro:2016wsq}
and use them in the context of multi-loop numerical unitarity.
More concretely, there are two components of a typical numerical
unitarity formulation that must be re-examined: (1) the generation of
a set of four-momenta satisfying momentum
conservation and on-shell conditions for the external kinematics, and
(2) the parameterization of the internal loop momenta, in particular
for the solution of the quadratic on-shell equations
and the construction of specific sets of surface terms. In this section
we will describe how these obstacles can indeed be overcome. Furthermore,
this will be achieved without needing to introduce any notion of complex numbers. In the
following we work over an arbitrary field which we denote by
$\mathbb{F}$, and which for all practical purposes can be considered to be
the rational numbers $\mathbb{Q}$.

\subsubsection{External Kinematics}\label{sec:FFExtKin}
The problem of generating a set of $\mathbb{F}$-valued external momenta
which satisfy on-shell conditions as well as
momentum conservation can be solved in a number of ways.
In this work, we choose to parameterize the external
kinematics based upon so-called ``momentum twistors''
\cite{Hodges:2009hk} as proposed in refs.~\cite{Badger:2012dp,Peraro:2016wsq},
where the reader can find further details.
In short, momentum twistors are particularly useful because they trivialize the on-shell
and the momentum conservation conditions.
By starting with an
$\mathbb{F}$-valued parameterization in twistor space, one obtains
 $\mathbb{F}$-valued spinors associated to each momentum.
If working with the usual Minkowski metric $\textrm{diag}\{+1,-1,-1,-1\}$,
the associated four-momenta and polarization vectors would involve explicit factors of $i$
and thus require an extension of $\mathbb{F}$.
This can be avoided by using the metric $\textrm{diag}\{+1,-1,+1,-1 \}$,
in which case the momenta and polarization vectors constructed from the
$\mathbb{F}$-valued spinors
are themselves $\mathbb{F}$-valued.\footnote{
In the standard formulae for gluonic polarization vectors
there is an explicit factor of $\sqrt{2}$:
\begin{equation}
  \epsilon^\mu_+ (p,\eta)=
  \frac{\langle \eta|\sigma^\mu|\left.\!\!p\right]}{\sqrt{2} \langle \eta p\rangle},
  \qquad \qquad
  \epsilon^\mu_- (p,\eta)=
  \frac{\langle p|\sigma^\mu | \left.\!\!\eta\right]}{\sqrt{2} \left[ p \eta\right]}.
  \label{eq:SpinorPolarizationVector}
\end{equation}
This is not an issue as it can be restored as a global factor after computation.
}
Avoiding the introduction of complex values offers a considerable speed
boost in a numerical implementation.
We also note that
a wisely chosen
parameterization such as that of \cite{Peraro:2016wsq} means that the
invariants and Gram-determinants are compact functions of the
parameters. This simplifies the functional dependence of the amplitude
on these parameters, making them particularly suitable for a future
reconstruction of the analytic expressions.

\subsubsection{On-shell Momenta}\label{sec:on-shellMom}

As discussed around eq.~\eqref{eq:CE}, in order
to compute integral
coefficients in a unitarity approach we need to generate loop momenta which satisfy a set of
conditions which set some propagators on-shell. 
For a two-loop calculation, this set of topology specific, 
kinematically-dependent quadratic conditions define an algebraic variety in the
6-dimensional space in which we embed the loop momenta.
A direct approach for finding a rational parameterization of this
variety, i.e., a parameterization in terms of a set of $\mathbb{F}$-valued parameters
that only uses the operations that are defined on a field, is non-trivial.
Instead we take inspiration from the fact
that the integrand is a rational function of irreducible scalar
products (ISPs), and the $\mu_{ij}$ variables which we shall define shortly.
We will see that in a set of
adapted coordinates it is trivial to generate loop momenta such that
the ISPs and $\mu_{ij}$ are $\mathbb{F}$-valued. Therefore, the integrand
evaluated on such a loop-momentum configuration will also be
$\mathbb{F}$-valued. We then represent the loop momenta in a phase-space dependent way,
circumventing the rational parameterization required when using the
standard 6-dimensional representation. We now give more details
about this procedure.

\begin{figure}[ht] 
   \includegraphics[scale=0.85]{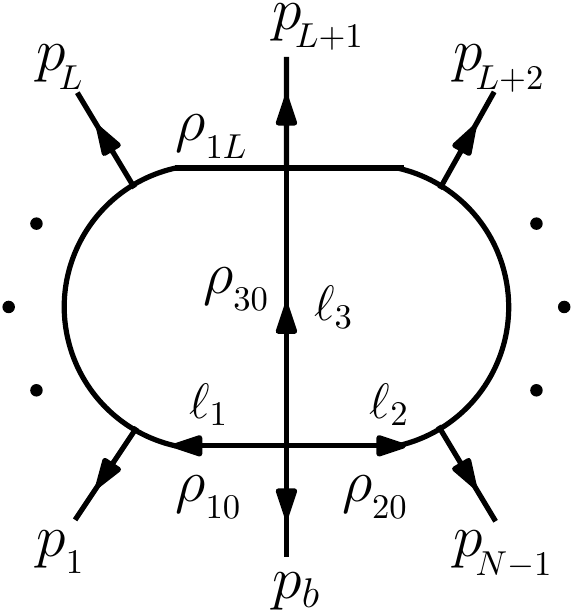}
   \caption{ Displayed are the conventions for assigning propagators in a
   two-loop diagram.} 
\label{fig:diagConventions} 
\end{figure}

We begin by parameterizing the loop momenta $\ell_l$
($l=1,2,3$ see fig. \ref{fig:diagConventions}) as
\cite{Ita:2015tya,Abreu:2017xsl} 
\begin{align} \ell_l &=
        \sum_{j \in B^p_l} v^j_l r^{l j} + \sum_{j \in B^t_l} v^j_l \alpha^{l
        j} + \sum_{i \in B^{ct}} \frac{n^i}{(n^i)^2}  \alpha^{l i} + \sum_{i \in B^{\epsilon}}
        n^i  \mu_l^{i}, \label{eq:AdaptedCoordinates} \\ 
        r^{l j}& = -\frac{1}{2} ( \rho_{lj} - (q_{lj})^2 - \rho_{l(j-1)} + (q_{l(j-1)})^2 )\,,  \\
        \mu_{ll} &
        = \rho_{l0} - \sum_{\nu = 0}^3\ell_l^\nu \ell_{l\,\nu}\,,
        \label{eq:MuRelation} 
\end{align} 
which are related by momentum conservation,
$\ell_1+\ell_2+\ell_3+p_b=0$. The vectors $q_{lj}$ are
linear combinations of the external momenta $p_i$.
The loop momenta are parameterized in terms of so-called
adapted coordinates: the inverse propagator variables
$\rho_{li}$, and the auxiliary variables $\alpha^{li}$ and $\mu^i_l$.
The variables $\mu^i_l$ are dependent and fixed by
(\ref{eq:MuRelation}). The vectors $n^i$ form an orthogonal basis
transverse to the scattering plane, i.e
$n^i\cdot p_j = 0$.  Labels in $B^\epsilon$ refer to directions beyond
four-dimensions and labels in $B^{ct}$ denote transverse directions
within four dimensions.
For each strand $l$ of the diagram we use a distinct basis of the scattering plane,
spanned by the vectors $v^i_l$, 
\begin{align} 
    v^i_l=(G_l)^{ij}p_j\,, \quad \mbox{with} \quad i,j\in B^p_l\cup B^t_l\,,
\end{align} 
where $(G_l)^{ij}$ is the inverse of the Gram matrix,
\begin{align} 
    (G_l)_{ij}=p_i\cdot p_j \quad \mbox{with} \quad i,j\in B^p_l\cup B^t_l\,.
\end{align} 
The index set $B^p_l$ labels the external momenta which leave the
strand $l$.  These momenta are completed with other independent
external momenta $p_i$, with $i\in B^t_l$, so as to span the whole
scattering plane.
This parameterization follows the conventions of
ref.~\cite{Abreu:2017xsl}, with the caveat that
the vectors spanning $B^{ct}$ are no longer normalized.

The inverse coordinate transformation is often useful and is given by
\begin{align} 
 \alpha^{li} &= p_i\cdot \ell_l \,,\quad i \in  B^t_l \,, \\
  \label{eq:CTExpression}
 \alpha^{li} &= n^i\cdot \ell_l \,,\quad i \in B^{ct} \,, \\
 \rho_{li} &= (\ell_l - q_{li})^2\,
\end{align} 
The on-shell variety is then defined by setting the propagator variables
$\rho_{li}$ to zero.  In $D$-dimensions the variables $\alpha^{li}$
form an independent complete set of coordinates on the variety,
corresponding to polynomials in momentum variables.  They are the
irreducible scalar products we already mentioned previously.

By considering the $\alpha^{li}$ as a set of independent coordinates and
taking them as $\mathbb{F}$-valued, the constraint
equations \eqref{eq:MuRelation} imply that the $\mu_{ij}$ are also $\mathbb{F}$-valued. We must now construct
an explicit representation of the loop momenta.
This is required, for instance, for the calculation of the tree
amplitudes in eq.~\eqref{eq:CE}. In the 4-dimensional
slice this is trivial to achieve with a standard cartesian
basis. However, if we were to also do this for the $(D-4)$-dimensional
space we would generically be required to take square roots.

We now present our solution to this problem in the context
of a two-loop calculation, but it trivially generalizes to
any loop order.
The main idea is to employ a different basis of the $(D-4)$-dimensional space
for each on-shell phase-space point. 
This is achieved by picking basis vectors which are linear combinations
of the $(D-4)$-dimensional components of the loop momenta.  Given the
two loop momenta,
\begin{equation}
    \ell_1=\left(\ell_{1,[4d]}, \vec\mu_1 \right)\,,\quad 
    \ell_2=\left(\ell_{2,[4d]}, \vec\mu_2 \right)\,,
\end{equation}
and their $(D-4)$-dimensional parts $\vec\mu_1$ and
$\vec\mu_2$, we construct the orthogonal basis vectors\footnote{Also at higher loops, one can construct orthogonal combinations through a modified Gram-Schmidt procedure.}
\begin{equation}
    \tilde{n}^1 = \left( 0_{[4d]} , \vec\mu_1 \right), \quad\quad 
    \tilde{n}^2 = \left( 0_{[4d]},\vec\mu_2   -  \frac{ \mu_{12}  }{\mu_{11} }\vec\mu_1 \right)\,,
\label{eq:eNormalization}
\end{equation}
with $\mu_{ij}=-\vec\mu_i\cdot\vec\mu_j$, here using the {\it Euclidean} scalar product.
We stress that the basis vectors $\tilde{n}^1$ and $\tilde{n}^2$
used to represent the $(D-4)$-dimensional space
are no longer of unit norm. For each on-shell point,
which corresponds to a different value of the $\mu_{ij}$,
this affects how we calculate the scalar
product between two vectors $w_a$ and $w_b$. Explicitly,
\begin{equation}
  w_a \cdot w_b = w_{a,[4d]} \cdot w_{b,[4d]} + w_a^5 w_b^5 \left(\mu_{11}\right) + w_a^6 w_b^6 \left(\mu_{22}-\frac{\mu_{12}^2}{\mu_{11}}\right).
\end{equation}
Such a scalar product then allows $\mathbb{F}$-valued representations
of loop momenta which satisfy the on-shell conditions for an arbitrary
topology. 

This momentum representation is almost all that is required to use a
Berends-Giele recursion~\cite{Berends:1987me} to calculate products of
tree amplitudes.  The remaining difficulty is in performing the sum
over helicities at each cut line, as this requires explicit
representations of the $D_s$-dimensional polarization vectors. To
remedy this, we choose to avoid constructing the polarization
states by trading the helicity sum for the insertion of a light-cone
projection operator (see e.g.~\cite{Ellis:2011cr}),
\begin{equation} P_l^{\mu \nu} = -g^{\mu \nu} + \frac{\ell_l^\mu \eta^\nu +
    \eta^\mu \ell_l^\nu}{\eta\cdot \ell_l}, \label{eq:LightConeProjector}
\end{equation} 
where $\eta$ is an arbitrary $\mathbb{F}$-valued light-like reference vector satisfying
$\eta \cdot \ell_l \ne 0$ and $\eta\cdot \tilde n = 0$. Note that in
all but one cut line per loop, the Ward identity allows us to drop the
second term of equation \eqref{eq:LightConeProjector}.
For the remaining cut line of each loop, we re-express the projector as a
sum of a direct product of vectors over $\mathbb{F}$.

We note that as the irreducible scalar products of
eq.~\eqref{eq:CTExpression} are expressed in terms of a set of
non-normalized vectors, this affects the representation of so-called
`traceless completion' surface terms \cite{Abreu:2017xsl},
i.e.~surface terms associated with variables
in $B^{ct}\cup B^{\eps}$. This
normalization now explicitly appears in the parameterization
of the loop momenta, see eq.~\eqref{eq:AdaptedCoordinates}.
For example, consider a traceless completion surface term
associated to the transverse vector $n^i, i \in B^{ct}$,
\begin{equation} 
\label{eq:AdjustedSurface}
    \frac{\alpha_i^2}{n_{i}^2} - \frac{\mu_{11}}{D-4}.
\end{equation} 
Only by including the factor of $n_{i}^2$ is the numerator insertion
\eqref{eq:AdjustedSurface} a surface term.

We finish this section by noting that this procedure applies both for
planar and non-planar cases, and is easily generalized to higher (and
lower) loops. In summary, using the steps described in Sections \ref{sec:FFExtKin} and \ref{sec:on-shellMom}, all contributions are
manifestly $\mathbb{F}$-valued, and as an added benefit we never needed
to introduce complex numbers.


\section{Implementation for Planar Five-Gluon Amplitudes}
\label{sec:Implementation}

We have implemented the techniques described in section
\ref{sec:Methods} in a C++ code. In this section, we first discuss the
implementation of the cut equations. Then we describe how these are solved
in practice to compute the master integral coefficients and, finally,
how we obtain the amplitude at a given kinematic point.

\subsection{Construction of Cut Equations}

\begin{figure}[!h] \begin{tikzpicture}[scale=1.2]
    \node at (2.5,0){\includegraphics[scale=0.4]{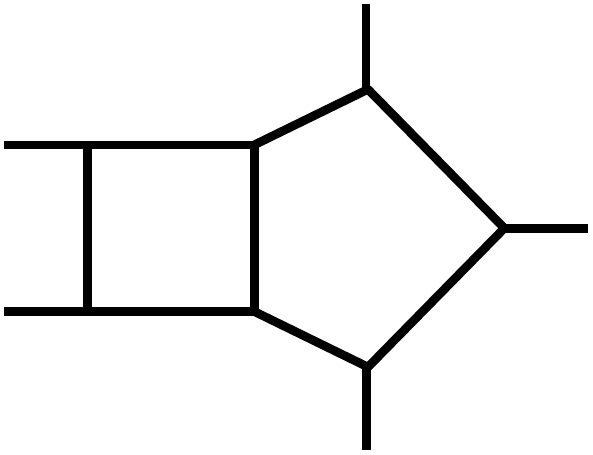}};
    \node at (5,0){\includegraphics[scale=0.4]{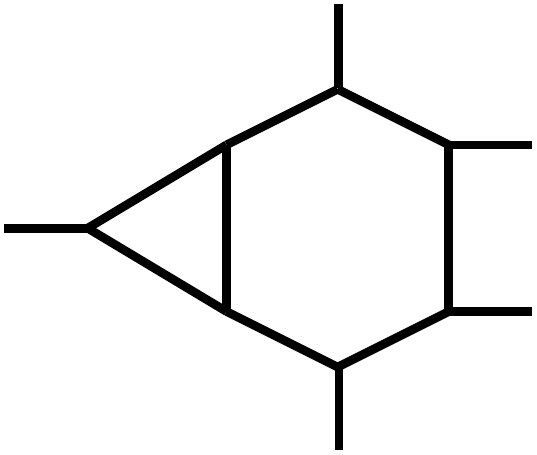}};
    \node at (7.5,0){\includegraphics[scale=0.4]{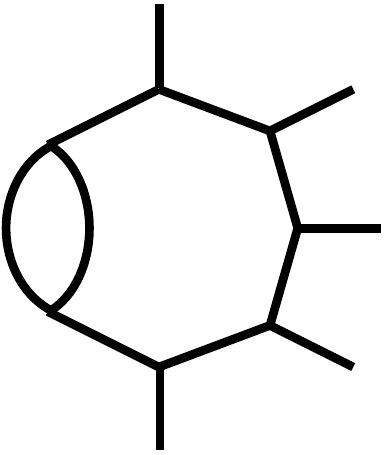}};
    \node at
    (0,-1.9){\includegraphics[scale=0.4]{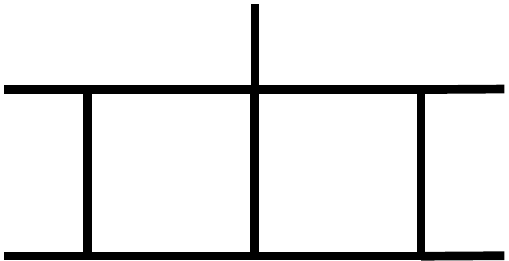}};
    \node at
    (2.5,-2.0){\includegraphics[scale=0.4]{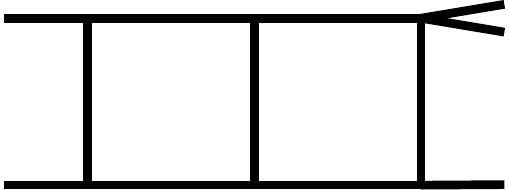}};
    \node at
    (5,-2.0){\includegraphics[scale=0.35]{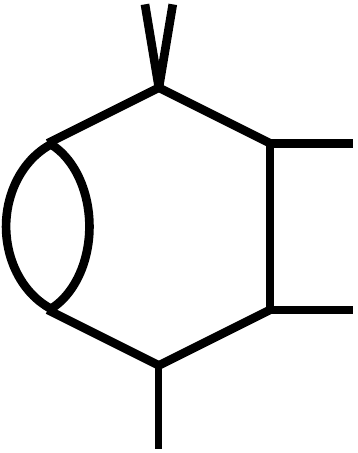}};
    \node at
    (7.5,-2.0){\includegraphics[scale=0.35]{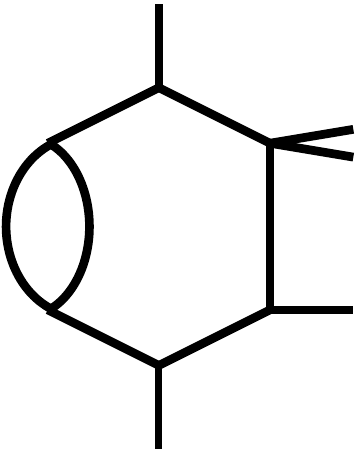}};
    \node at
    (10,-2.0){\includegraphics[scale=0.35]{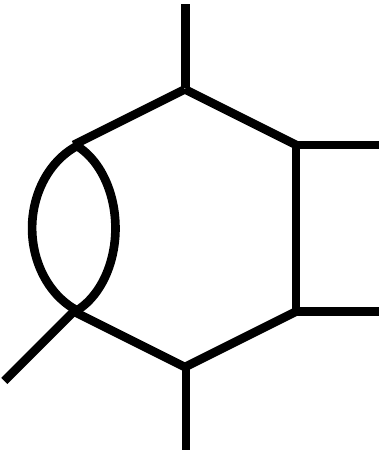}};
    \node at
    (0,-3.6){\includegraphics[scale=0.35]{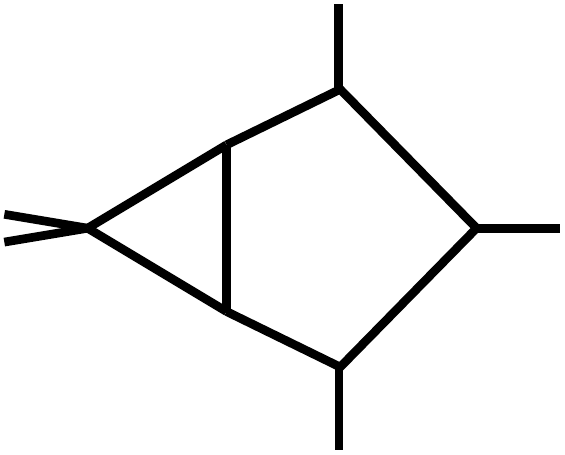}};
    \node at
    (2.5,-3.6){\includegraphics[scale=0.35]{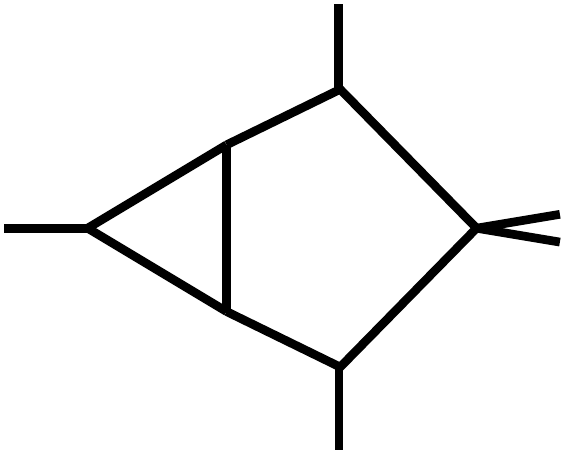}};
    \node at
    (5,-3.6){\includegraphics[scale=0.35]{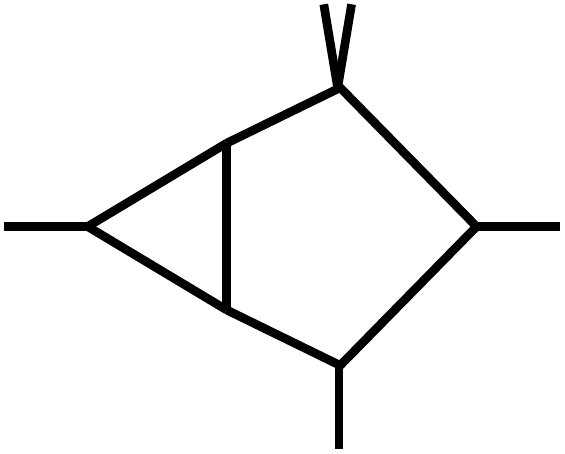}};
    \node at
    (7.5,-3.6){\includegraphics[scale=0.35]{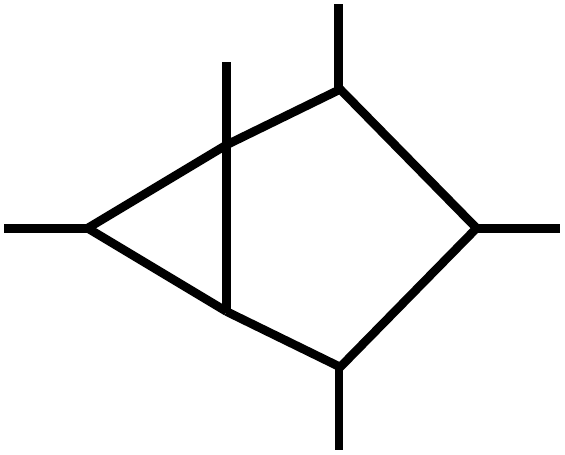}};
    \node at
    (10,-3.6){\includegraphics[scale=0.4]{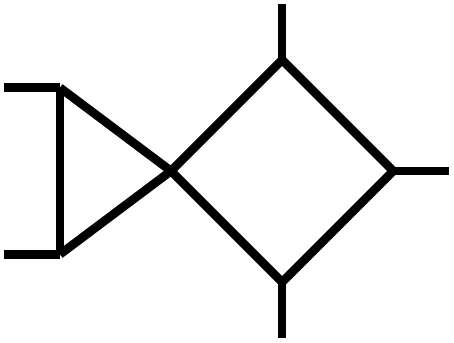}};
    \node at
    (0,-5.5){\includegraphics[scale=0.35]{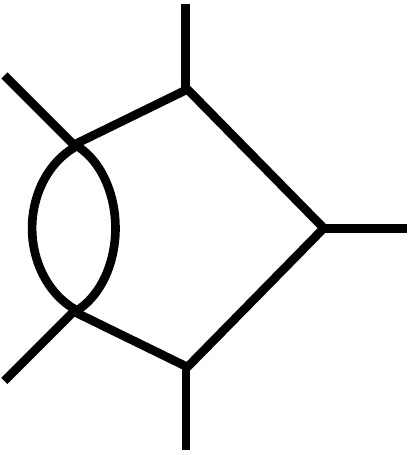}};
    \node at
    (2,-5.5){\includegraphics[scale=0.35]{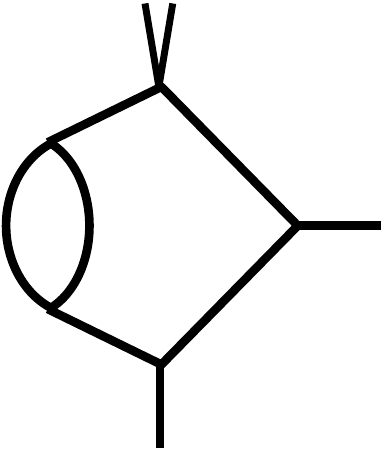}};
    \node at
    (4,-5.5){\includegraphics[scale=0.35]{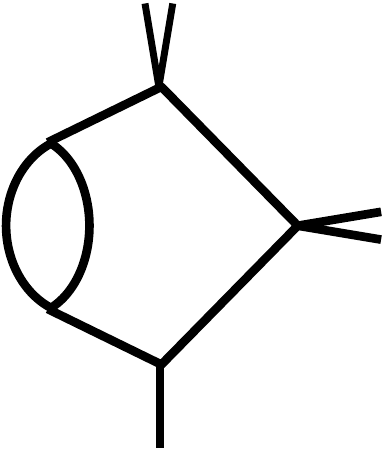}};
    \node at
    (6,-5.5){\includegraphics[scale=0.35]{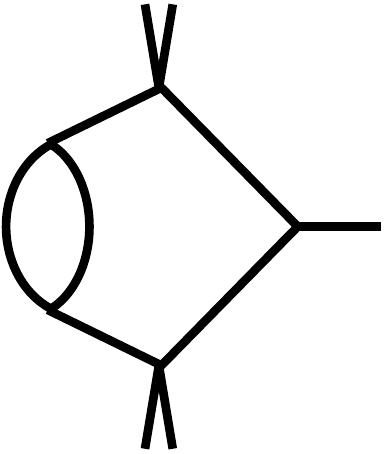}};
    \node at
    (8,-5.5){\includegraphics[scale=0.35]{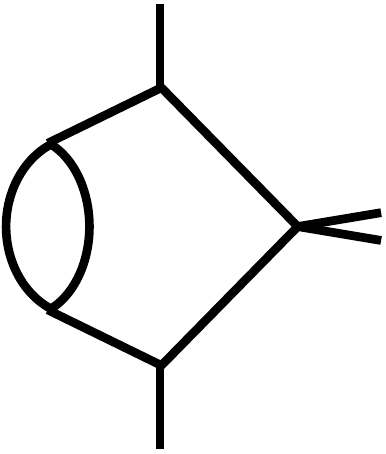}};
    \node at
    (10,-5.5){\includegraphics[scale=0.35]{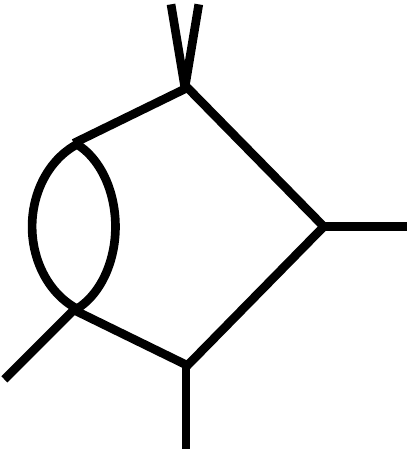}};
    \node at
    (0,-7){\includegraphics[scale=0.35]{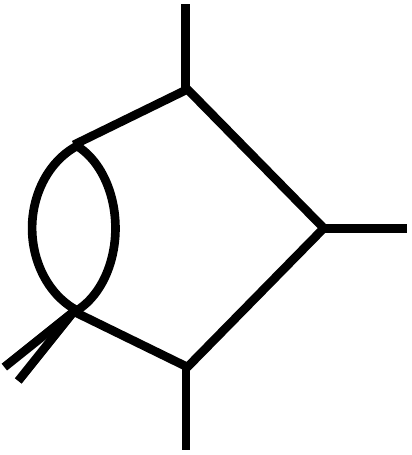}};
    \node at
    (2,-7){\includegraphics[scale=0.35]{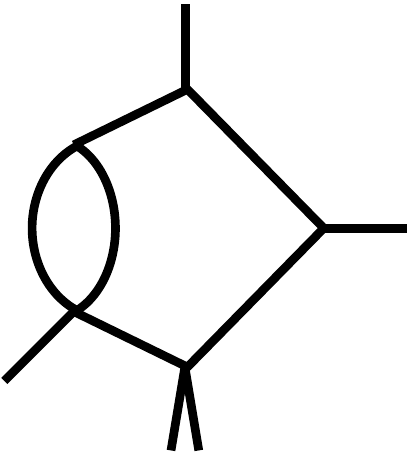}};
    \node at
    (4,-7){\includegraphics[scale=0.35]{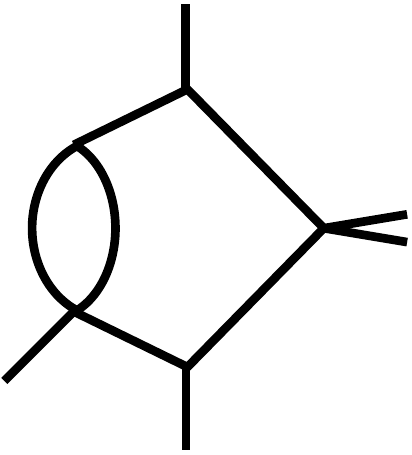}};
    \node at
    (6,-7){\includegraphics[scale=0.4]{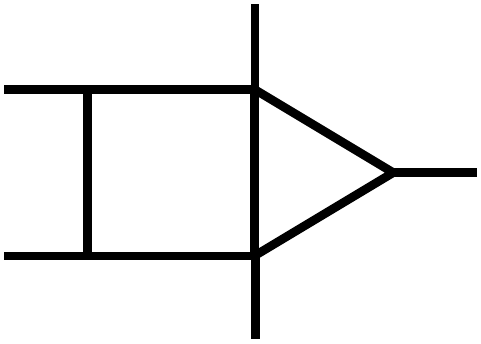}};
    \node at
    (8,-6.9){\includegraphics[scale=0.4]{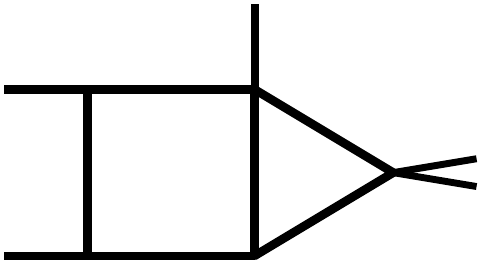}};
    \node at
    (10,-6.9){\includegraphics[scale=0.4]{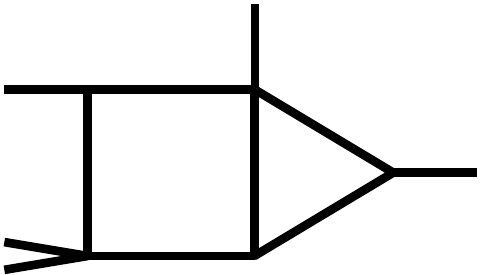}};
    \node at
    (0,-8.4){\includegraphics[scale=0.4]{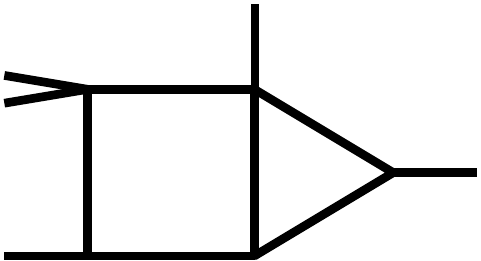}};
    \node at
    (2,-8.4){\includegraphics[scale=0.4]{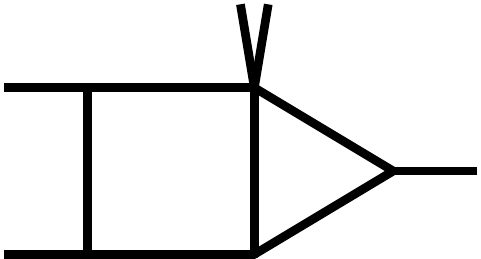}};
    \node at
    (4,-8.5){\includegraphics[scale=0.4]{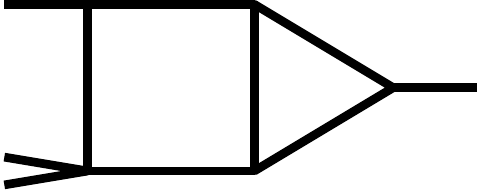}};
    \node at
    (6,-8.5){\includegraphics[scale=0.4]{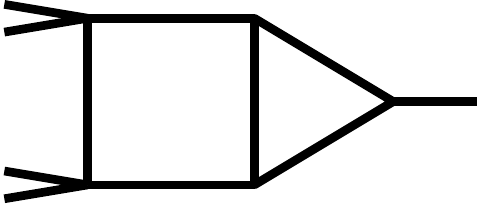}};
    \node at
    (8,-8.5){\includegraphics[scale=0.4]{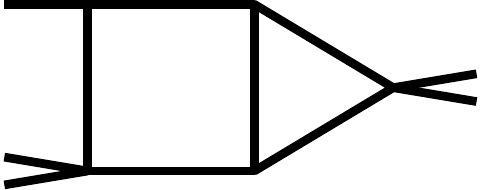}};
    \node at
    (10,-8.5){\includegraphics[scale=0.4]{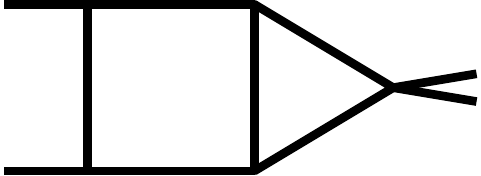}};
    \node at
    (3,-9.8){\includegraphics[scale=0.4]{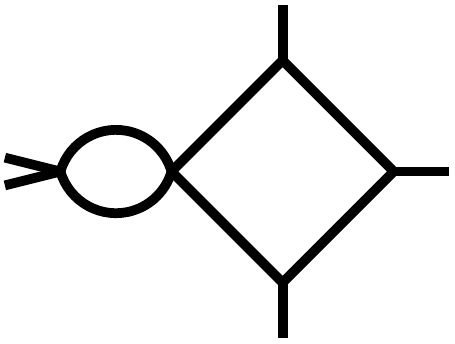}};
    \node at
    (5,-9.8){\includegraphics[scale=0.35]{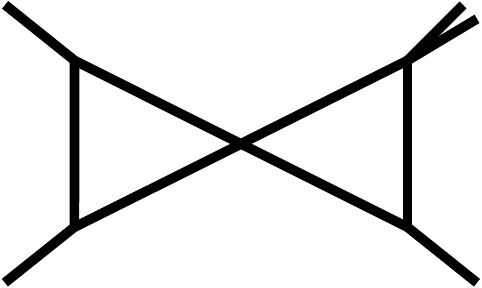}};
    \node at
    (7,-9.8){\includegraphics[scale=0.35]{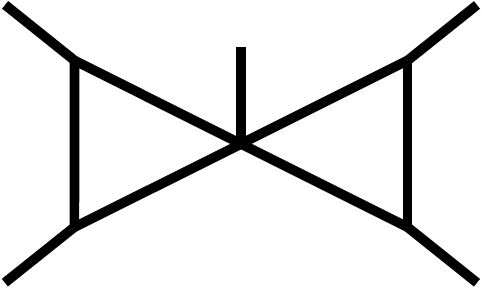}};
    \node at
    (0,-11.5){\includegraphics[scale=0.45]{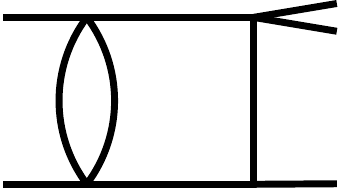}};
    \node at
    (2,-11.5){\includegraphics[scale=0.45]{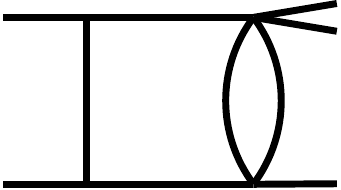}};
    \node at
    (4,-11.5){\includegraphics[scale=0.45]{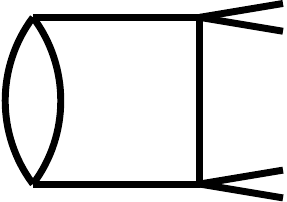}};
    \node at
    (6,-11.5){\includegraphics[scale=0.45]{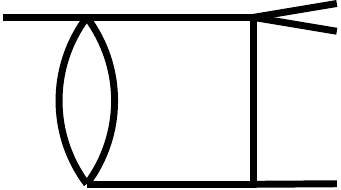}};
    \node at
    (8,-11.5){\includegraphics[scale=0.45]{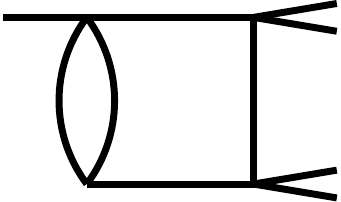}};
    \node at
    (10,-11.5){\includegraphics[scale=0.45]{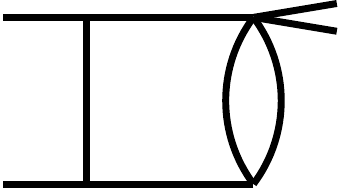}};
    \node at
    (0,-12.7){\includegraphics[scale=0.45]{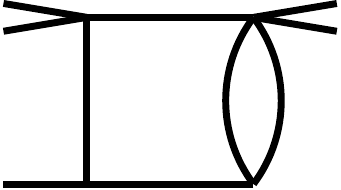}};
    \node at
    (2,-12.7){\includegraphics[scale=0.45]{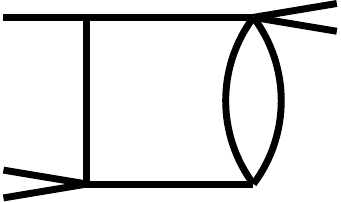}};
    \node at
    (4,-12.7){\includegraphics[scale=0.45]{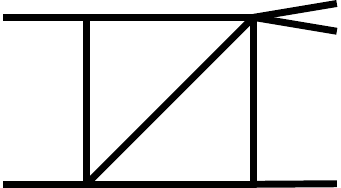}};
    \node at
    (6,-12.7){\includegraphics[scale=0.45]{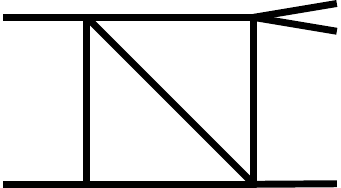}};
    \node at
    (8,-12.6){\includegraphics[scale=0.45]{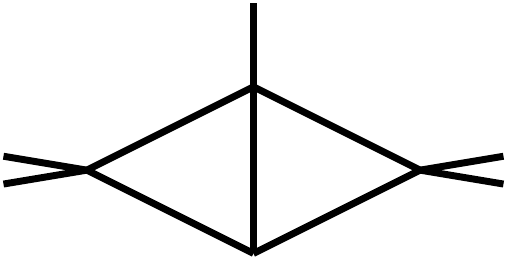}};
    \node at
    (10,-12.75){\includegraphics[scale=0.45]{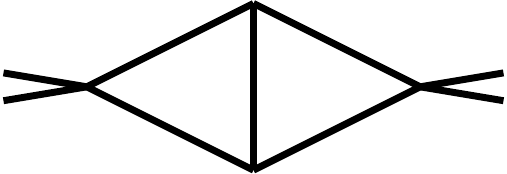}};
    \node at
    (0,-13.8){\includegraphics[scale=0.45]{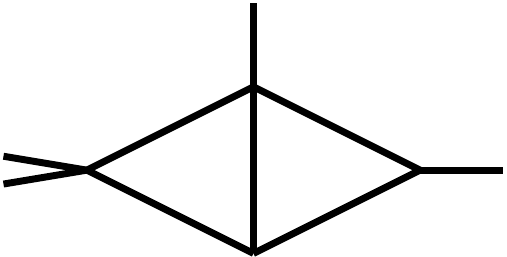}};
    \node at
    (2,-13.8){\includegraphics[scale=0.45]{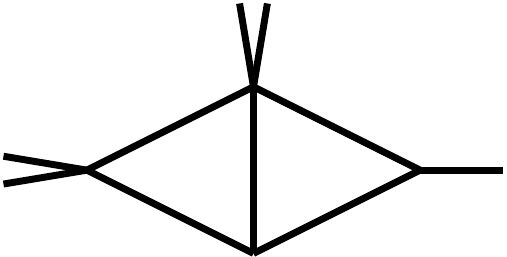}};
    \node at
    (4,-13.8){\includegraphics[scale=0.45]{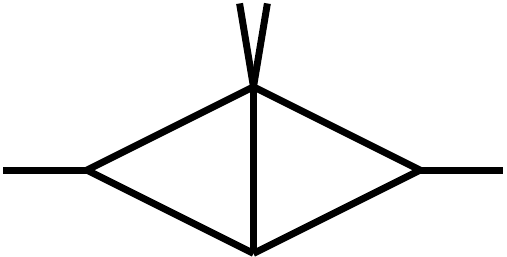}};
    \node at
    (6,-13.95){\includegraphics[scale=0.4]{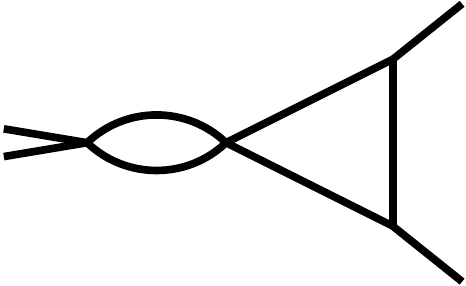}};
    \node at
    (8,-13.95){\includegraphics[scale=0.4]{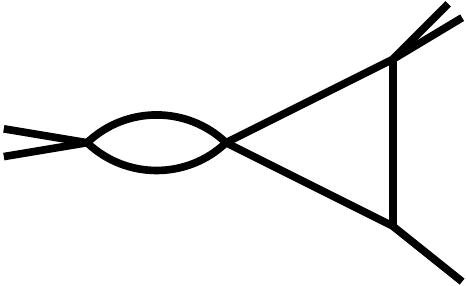}};
    \node at
    (10,-13.95){\includegraphics[scale=0.4]{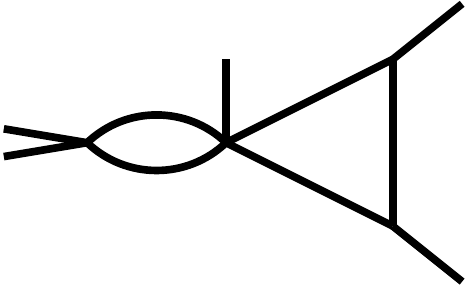}};
    \node at
    (0,-15.5){\includegraphics[scale=0.5]{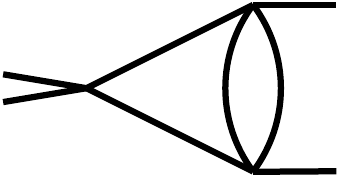}};
    \node at
    (1.8,-15.5){\includegraphics[scale=0.5]{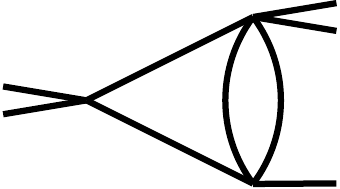}};
    \node at
    (3.6,-15.5){\includegraphics[scale=0.5]{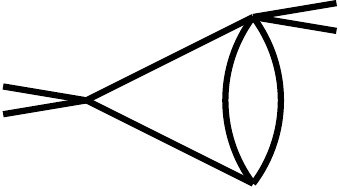}};
    \node at
    (5.4,-15.5){\includegraphics[scale=0.5]{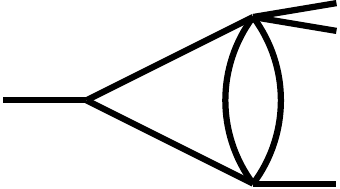}};
    \node at
    (7.2,-15.5){\includegraphics[scale=0.5]{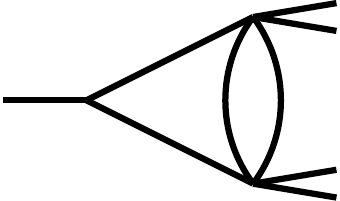}};
    \node at
    (9,-15.55){\includegraphics[scale=0.5]{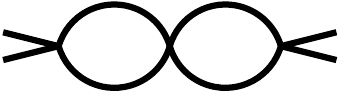}};
    \node at
    (10.8,-15.5){\includegraphics[scale=0.5]{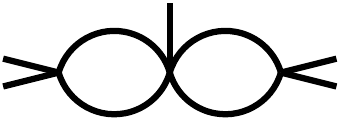}};
    \node at
    (5.25,-17){\includegraphics[scale=0.6]{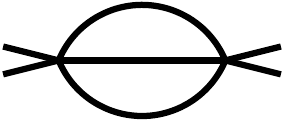}};
\end{tikzpicture} \caption{Hierarchy of propagator structures for two-loop five-point gluon scattering. Only topologically inequivalent structures are shown.}
\label{fig:PropagatorStructures}
\end{figure}

We first construct the full set of propagator structures that appear
in the problem (see figure \ref{fig:PropagatorStructures}). This is
achieved by generating all cut diagrams for the full color
process using QGRAF \cite{Nogueira:1991ex} and then color decomposing
them according to \cite{Ochirov:2016ewn} in \Mathematica{}. By taking
the leading-color limit of this decomposition and extracting the
coefficient of a given trace, we build the set of propagator structures
relevant for the color ordered amplitude, which we then organize
hierarchically. This is then passed on to a C++ code.
%

The next step in constructing the cut equations is
the parameterization of the integrand of the amplitude
in terms of master integrands and surface terms.
In order to solve the syzygy equations described in section \ref{sec:IBP} for
each topology in fig.~\ref{fig:PropagatorStructures} we use the
package {\sc Singular}.  For these
computations we find that the `slimgb' algorithm
\cite{brickenstein2010slimgb} is much faster than the classical
Buchberger algorithm.
 The `slimgb' algorithm uses `weighted lengths' of polynomials to make choices in key steps of the computation, in order to keep the size of intermediate results small.
For complicated topologies we further speed up the computation by imposing a
degree bound in {\sc Singular}, which restricts the maximum allowed polynomial
degree in internal computations. Another significant speed-up is achieved by
promoting parameters of the polynomial ring, which are external kinematic
invariants, to variables of the polynomial ring, so that the solutions are
restricted to have polynomial (instead of rational) dependence on these
kinematic invariants. In practice this restriction does not prevent us from
finding the solutions that lead to a complete set of IBP
relations. For one of the most complicated topologies involved in our
calculation, namely the pentabox shown on the top left of
fig.~\ref{fig:PropagatorStructures}, our method succeeds in finding fully
off-shell IBP-generating vectors, with analytic dependence on all external
kinematic invariants, in under a second.
The computation used only one CPU core on a modern laptop computer.
All surface terms have been validated with \texttt{Fire}
\cite{Smirnov:2008iw,Smirnov:2014hma} on a fixed numerical kinematic
point.

We note that a fast evaluation of the surface terms is required for an
efficient implementation.
To this end
we express the surface terms as IBP-generating vector components multiplied by 
derivatives of the irreducible scalar products, as well as the total 
divergences of the vectors multiplied by irreducible scalar products.
Furthermore, to improve both evaluation time and compilation time of
the IBP vector components, we found it useful to employ the facilities
provided by FORM \cite{Vermaseren:2000nd} for the simplification of
multivariate polynomials~\cite{Kuipers:2012rf,Ruijl:2014spa}.

Given the parameterization of the integrand with surface terms, any
linearly-independent tensor insertion can be used as a master
integral. As such, it is trivial for us to change basis of master integrals,
by filling the master space of each topology with any set of integrals
independent under the IBP relations. This freedom was especially useful
during testing, where we could make use of finite integrals to control
the $\epsilon$ structure.

In order to numerically calculate the appropriate products of tree
amplitudes necessary to reconstruct the integrand in our algorithm, we
implement a Berends-Giele recursion~\cite{Berends:1987me} tailored to directly
compute multi-loop cuts using $D$-dimensional momenta and states living in $D_s$
dimensions. This is both a high-performance
and flexible choice, as changing the field content requires only
implementing new Feynman rules. The setup is particularly useful for
our numerical computations in dimensional regularization, as it is
straightforward to evaluate the products of tree amplitudes at different
values of $D_s$ in order to reconstruct the functional dependence on
the parameter, in a way that automatically recognizes if a given two-loop cut has
a linear or quadratic dependence on $D_s$. Caching for multiple $D_s$ values is
built-in.
External momenta are taken to live in 4 dimensions, as do the
associated gluonic polarization states.  The $D$-dimensional loop momenta
are represented in 6-dimensions, as described in detail in
section \ref{sec:FiniteFields}.

\subsection{Amplitude Evaluation}

Having constructed the cut equations (\ref{eq:CE}) as described in the
previous section, we now solve them for the integral coefficients at
fixed values of the kinematics. Both equation (\ref{eq:CE}) and its
hierarchically subtracted analogue describe linear systems in the
ansatz coefficients. For fixed values of $D$ and $D_s$, we evaluate
the equations numerically over enough randomly chosen on-shell momenta
configurations 
to form a linear system that constrains
the coefficients. We then solve this
system for the coefficients using standard PLU 
factorization and back substitution.

In order to reconstruct the $D$ and $D_s$ dependence of the coefficients
we first sample $D_s$ over three distinct values \cite{Abreu:2017xsl}
in a generalization of \cite{Giele:2008ve}, allowing us to easily
implement both the FDH and HV flavors of dimensional
regularization. As $D_s$ is restricted to be greater than or equal to
the embedding space of the loop momenta (6 for a two-loop calculation)
we pick $D_s = {6,7,8}$. The values
of $D$ are chosen randomly. The $D$-dependence is obtained by
repeating the computation for sufficient (a priori unknown) values of
$D$ from which the rational dependence is
reconstructed~\cite{vonManteuffel:2014ixa, Peraro:2016wsq} using
Thiele's formula \cite{Peraro:2016wsq, abramowitz1964handbook}. After
evaluation at a single phase-space point, the exact denominator as
well as rank of the numerator function can be stored, allowing the use
of simple polynomial inversion techniques and less sampling.

In practice, implementing our approach over the rational numbers
$\mathbb{Q}$ will result in a slow calculation. However, the final
master integral coefficients are strongly constrained by physical
properties and it is expected that their resulting form will be
compact. We thus follow the approach outlined in
\cite{vonManteuffel:2014ixa, Peraro:2016wsq}, using finite fields
$\mathbb{F}_p$. We implement the algorithm over the finite fields
provided by Givaro \cite{Givaro}. We use various cardinalities of
order $2^{30}$, implementing Barrett reduction
\cite{Barrett1987,HoevenLQ14} in order to improve the speed of finite
field multiplication.  For a given kinematic point in $\mathbb{Q}$, we
perform the computation in a sufficient number of finite fields to
apply a rational reconstruction algorithm, also provided by Givaro.

\begin{figure}[!h]
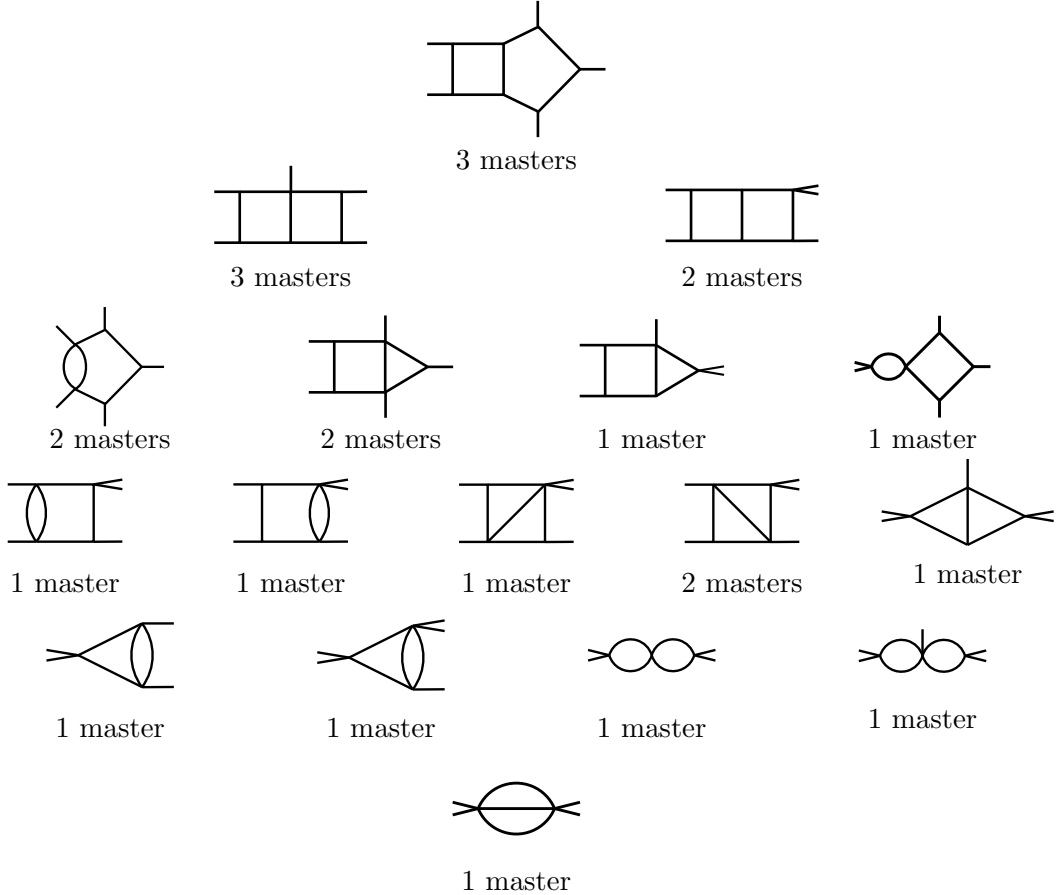
 \begin{tikzpicture}[scale=1.2]
    \node at (5,0){\includegraphics[scale=0.4]{pictures/BoxPentagon}};
    \node at (5,-1){3 masters};
    \node at
    (2.5,-1.5){\includegraphics[scale=0.4]{pictures/BoxBoxSG}};
    \node at (2.5,-2.3){3 masters};
    \node at
    (7.5,-1.6){\includegraphics[scale=0.4]{pictures/BoxBox}};
    \node at (7.5,-2.3){2 masters};
    \node at
    (0.5,-3.3){\includegraphics[scale=0.35]{pictures/BubblePentagonG}};
    \node at (0.5,-4.1){2 masters};
    \node at
    (3.5,-3.3){\includegraphics[scale=0.4]{pictures/BoxTriangleG}};
    \node at (3.5,-4.1){2 masters};
    \node at
    (6.5,-3.2){\includegraphics[scale=0.4]{pictures/BoxTriangleSG}};
    \node at (6.5,-4.1){1 master};
    \node at
    (9.5,-3.3){\includegraphics[scale=0.4]{pictures/BoxBubble1LS}};
    \node at (9.5,-4.1){1 master};
    \node at
    (0,-4.9){\includegraphics[scale=0.45]{pictures/BoxBubbleGBox}};
    \node at (0,-5.7){1 master};
    \node at
    (2.5,-4.9){\includegraphics[scale=0.45]{pictures/BoxBubbleGBub}};
    \node at (2.5,-5.7){1 master};
    \node at
    (5,-4.9){\includegraphics[scale=0.45]{pictures/TriangleTriangleGDiag}};
    \node at (5,-5.7){1 master};
    \node at
    (7.5,-4.9){\includegraphics[scale=0.45]{pictures/TriangleTriangleGOffDiag}};
    \node at (7.5,-5.7){2 masters};
    \node at
    (10,-4.8){\includegraphics[scale=0.45]{pictures/TriangleTriangleSG}};
    \node at (10,-5.6){1 master};
    \node at
    (0.5,-6.5){\includegraphics[scale=0.5]{pictures/BubbleTriangleG1Mass}};
    \node at (0.5,-7.3){1 master};
    \node at
    (3.5,-6.5){\includegraphics[scale=0.5]{pictures/BubbleTriangleG2Mass}};
    \node at (3.5,-7.3){1 master};
    \node at
    (6.5,-6.5){\includegraphics[scale=0.5]{pictures/BubbleBubble1LS}};
    \node at (6.5,-7.3){1 master};
    \node at
    (9.5,-6.45){\includegraphics[scale=0.5]{pictures/BubbleBubble1LS_2}};
    \node at (9.5,-7.2){1 master};
    \node at
    (5,-8.2){\includegraphics[scale=0.6]{pictures/Sunrise}};
    \node at (5,-9){1 master};
\end{tikzpicture} \caption{Propagator structures with master integrals}
\label{fig_master_int}
\end{figure}

In order to obtain an $\epsilon$ expansion for the amplitudes, we
combine the coefficients with $\epsilon$ expansions of the appropriate
master integrals. We list the topologies with master integrals in
figure \ref{fig_master_int}. For the five-point master integrals
we choose the basis of \cite{Papadopoulos:2015jft}
in order to make use of the publicly available implementation distributed with the paper.
For lower point integrals, we implemented the analytic expressions
provided in \cite{Gehrmann:2000zt}.
In the case of factorizable topologies we choose the
scalar integral as a master and calculate them independently. In order
to evaluate the necessary multiple polylogarithms we use the
implementation found in GiNaC~\cite{Vollinga:2004sn}, which can be
tuned to the desired precision. All integrals have been independently
numerically validated with \texttt{Fiesta}~4~\cite{Smirnov:2015mct}.
We note that the choice~\cite{Gehrmann:2000zt} of an integral with a squared central
propagator for the second slashed-box master is important in order to
avoid spurious weight-five contributions in intermediate stages. 

\section{Results for Helicity Amplitudes}
\label{sec:Results}

As an illustration of the implementation of our approach, we present
numerical results for the four independent helicity
configurations. We evaluate them at the Euclidean phase-space point\footnote{The
reason for the explicit factor of $i$ in the third component of the four-vectors is
that we write them here for the standard $\textrm{diag}(+,-,-,-)$ metric. In our implementation
we use the metric $\textrm{diag}(+,-,+,-)$, see section \ref{sec:FFExtKin}, in which
case all the components are real.}
\begin{align}\begin{split}
  p_1 &= \left(\frac{1}{2}, \frac{1}{16}, \frac{i}{16}, \frac{1}{2}\right), \quad
  p_2 = \left( -\frac{1}{2}, 0, 0, \frac{1}{2} \right),\quad
  p_3 = \left(\frac{9}{2}, -\frac{9}{2}, \frac{7i}{2}, \frac{7}{2}\right), \\
  p_4 &= \left(-\frac{23}{4}, \frac{61}{16}, -\frac{131i}{16}, -\frac{37}{4}\right), \quad
  p_5 = \left(\frac{5}{4}, \frac{5}{8}, \frac{37i}{8}, \frac{19}{4}\right), 
  \label{eq:EvalPoint}
\end{split}\end{align}
which corresponds to the set of invariants (we write $s_{ij}=(p_i+p_j)^2$)
\begin{equation}
  s_{12} = -1, \quad s_{23} = -8, \quad s_{34} = -10, \quad s_{45} = -7, \quad s_{51} = -3,
\end{equation}
and set the dimensional regularization scale~$\mu$ to 1.
The normalization of the results we present is fixed by the expansion in eq.~\eqref{eq:bareAmp}, and they are all given in the HV scheme, $D_s=D=4-2\epsilon$. Since our coefficients are exact rational numbers and we have analytic expressions for the master integrals~\cite{Papadopoulos:2015jft,Gehrmann:2000zt}, the precision of the results we present is only limited by the number of digits we ask from GiNaC~\cite{Vollinga:2004sn} when evaluating the master integrals, which can be arbitrarily increased.

The results for the all-plus and single-minus helicity amplitudes,
$\mathcal{A}^{(2)}(1^+, 2^+, 3^+, 4^+, 5^+)$ and $\mathcal{A}^{(2)}(1^-, 2^+, 3^+, 4^+, 5^+)$,
are given in table \ref{tab:Results}. Both helicity configurations vanish at tree-level,
which implies they are finite at one-loop and start at order $\eps^{-2}$ at two-loops.
We present our two-loop results normalized to $\mathcal{A}^{(1)}(\eps=0)$,
the finite one-loop amplitude truncated to leading order in $\epsilon$.
The results for the two independent MHV amplitudes,
the split $\mathcal{A}^{(2)}(1^-, 2^-, 3^+, 4^+, 5^+)$ and the alternating
$\mathcal{A}^{(2)}(1^-, 2^+, 3^-, 4^+, 5^+)$ helicity configurations,
are given in table \ref{tab:ResultsMHV}.
In this case, we normalize the results to $\mathcal{A}^{(0)}$,
 the corresponding tree-level amplitude.
As expected, the first two leading poles are helicity independent.

\begin{table}[htb]
\begin{tabular}{|c|c|c|c|}
  \hline
  $\mathcal{A}^{(2)}/\left(\mathcal{A}^{(1)}(\eps=0)\right)$ &\, $\epsilon^{-2}$\, & $\epsilon^{-1}$ & $\epsilon^{0}$ \\ \hline
  \, $(1^+, 2^+, 3^+, 4^+, 5^+)\, $\, &\,-5.000000000\, & \, -3.8931790255\, & 5.9810885816\, \\ \hline
  \, $(1^-, 2^+, 3^+, 4^+, 5^+)\, $\, &\,-5.000000000\, &\, -16.322002103\, &\, -10.383813287 \, \\ 
  \hline
\end{tabular}
\caption{Numeric results truncated to 10 significant figures for the
two-loop all-plus and single-minus helicity amplitudes, normalized to
the finite one-loop amplitudes truncated to leading order in
$\epsilon$, at the kinematic point of eq.~\eqref{eq:EvalPoint}.}
\label{tab:Results}
\end{table}

\begin{table}[htb]
\begin{tabular}{|c|c|c|c|c|c|}
  \hline
  $\mathcal{A}^{(2)}/\mathcal{A}^{(0)}$ & $\epsilon^{-4}$ & $\epsilon^{-3}$ & $\epsilon^{-2}$ & $\epsilon^{-1}$ & $\epsilon^{0}$ \\ \hline
  $\, (1^-, 2^-, 3^+, 4^+, 5^+)\, $ & \,12.5000000\,&\, 25.46246919\, &\, -1152.843107\, &\, -4072.938337\, &\, -3637.249567\, \\ \hline
  $\, (1^-, 2^+, 3^-, 4^+, 5^+)\, $ &\,12.5000000\, &\, 25.46246919\, &\, -6.121629624\, &\, -90.22184215\, &\, -115.7836685\, \\ 
  \hline
\end{tabular}
\caption{Numeric results truncated to 10 significant figures for the two-loop split and alternating MHV amplitudes, normalized to the tree level, at the kinematic point of eq.~\eqref{eq:EvalPoint}.}
\label{tab:ResultsMHV}
\end{table}

To validate our results, we first reproduce the universal ultraviolet/infrared
pole structure of the amplitudes \cite{Catani:1998bh}, which 
we summarize in appendix \ref{sec:IR} in the context of two-loop
five-gluon amplitudes in the leading-color approximation.
Computing the prediction for the pole structure requires
the five-point one-loop amplitudes up to order $\epsilon$
for all four independent helicity configurations.
For this, we used our own implementation of numerical unitarity at one loop,
which we checked up to order $\eps^0$ against results from
\BlackHat~\cite{Berger:2008sj}. Within our check,
the precision of the numerical results we obtain for the poles
is limited by the fact that we use \texttt{Fiesta}~4~\cite{Smirnov:2015mct}
for the one-loop pentagon integral.
Using our code, we confirm the published
  results for the all-plus helicity amplitudes \cite{Badger:2013gxa,Gehrmann:2015bfy,Dunbar:2016aux}.
We also validate the results of~\cite{Badger:2017jhb} which appeared during the final stages
 of the preparation of this article.

 Finally, we note that although we only present results for the four independent helicity configurations, we have also verified the pole structure for the other helicity configurations obtained by parity conjugation or permutation of external legs.
Since our calculation is based on a numerical setup, these amount to independent calculations that give us an internal consistency check of our implementation.


\section{Conclusion}
\label{sec:Conclusion}

The main result of this paper is a numerical implementation
of two-loop five-gluon amplitudes in the leading-color
approximation, for any helicity configuration. The calculation
is done in the multi-loop numerical unitarity approach,
which we extended for use with finite-field arithmetics.
Using a new approach to the generation of IBP relations,
we obtain a complete decomposition of the amplitude in terms
of master integrals.
Because of the extension to finite-field arithmetics, the
corresponding coefficients computed from unitarity cuts are exact.
To illustrate our calculation, we presented
the results at a given kinematic point, for which
we used the available master integrals \cite{Gehrmann:2000zt,Papadopoulos:2015jft}.
Our implementation was validated by checking the universal pole structure \cite{Catani:1998bh},
and confirms all results available in the literature
\cite{Badger:2013gxa,Gehrmann:2015bfy,Dunbar:2016aux,Badger:2017jhb}.

The high polynomial degree and the many scales
associated with multi-loop computations with several external particles
make it challenging to
set up accurate numerical approaches.
While analytic results are stable, the complexity of intermediate
computational steps often makes them difficult to obtain for
challenging amplitudes of that kind.
Here, we chose an alternative approach which removes the strict separation
of numerical and analytic perspectives and keeps the best features of each.
We employed exact arithmetics as previously
explored in the context of generalized unitarity in ref.~\cite{Peraro:2016wsq}.
Our approach is universal and is obtained after
refinements of the method introduced in \cite{Abreu:2017xsl, Abreu:2017idw}.
The exact approach is valuable for a number of reasons:
the validation of results is simplified because the results are exact,
which allows us to evaluate the amplitudes in singular limits,
and, in addition, it paves the way for the
functional reconstruction of their analytic form.
This is left for future work.

The potential to generalize to other processes is significant, based on our
experience with the present computation. The approach is general and requires
little spectrum-dependent work.
In particular, the parameterization of the integrand of the amplitude
in terms of master integrals and surface terms, which allows to achieve
a full reduction to master integrals, only depends on the kinematics
of the process and the power-counting of the theory.
The current setup handles five kinematic scales and the generalization to
other massless particles is straightforward.
Based on the
present computation, we also believe that general five-scale processes are attainable
and that adding further mass scales will be achievable in the near
future.  Finally, extensions to non-planar amplitudes appear  well within reach.

Given the importance of analytic results for inspiring new methods and for
identifying hidden symmetry principles, it will be exciting to learn about the
analytic form of the integral coefficients. With the presented methods, we are
well set up to explore this direction.

\begin{acknowledgments}
We thank Z.~Bern, M.~Jaquier, D.~A.~Kosower,
H.~Sch\"onemann and V. Sotnikov for helpful discussions. We thank
M.~Jaquier for early participation in this work.
We thank C.~Duhr for the use of his \Mathematica{} package \texttt{PolyLogTools}.
The work
of S.A., F.F.C., and B.P. is supported by the Alexander von Humboldt Foundation, in the
framework of the Sofja Kovalevskaja Award 2014, endowed by the German Federal
Ministry of Education and Research. 
H.I.'s work is supported by the Juniorprofessor Program of Ministry of Science, 
Research and the Arts of the state of Baden-W\"urttemberg, Germany.
The work of M.Z. is supported by the U.S. Department of Energy under Award Number
DE-{S}C0009937.
This work was performed on the bwUniCluster funded by the Ministry of Science,
Research and the Arts Baden-W\"urttemberg and the Universities of the State of
Baden-W\"urttemberg, Germany, within the framework program bwHP.  
This research is supported by the Munich Institute for Astro- and Particle
Physics (MIAPP) of the DFG cluster of excellence `Origin and Structure of the
Universe'.
\end{acknowledgments}

\appendix

\section{Divergence Structure of Two-Loop Gluon Amplitudes}
\label{sec:IR}

We present results for two-loop gluonic amplitudes with generic helicity configurations.
These amplitudes have both infrared and ultraviolet poles. The latter are removed through
renormalization of the amplitude. After renormalization,
the infrared poles are predicted from lower-order results through
a general formula~\cite{Catani:1998bh}. Here, we briefly summarize the procedure of
renormalization and the calculation of the infrared poles for the amplitudes we compute in
this paper. Reproducing the expected pole structure described in this section
is an important check of our results.

\subsection{Renormalization of Leading-Color Two-Loop Gluonic Amplitudes}

Renormalization of the amplitude is performed in the
$\overline{\text{MS}}$ scheme. It is implemented by replacing the bare
coupling by the renormalized one, denoted $\alpha_S$, in
eq.~\eqref{eq:bareAmp}.  The bare and renormalized couplings are
related through
\begin{equation}
    \alpha_0\mu_0^{2\epsilon}S_{\epsilon}
  =\alpha_s\mu^{2\epsilon}\left(
  1-\frac{\beta_0}{\epsilon}\frac{\alpha_s}{4\pi}
  +\left(\frac{\beta_0^2}{\epsilon^2}-\frac{\beta_1}{2\epsilon}\right)
  \left(\frac{\alpha_s}{4\pi}\right)^2+\mathcal{O}\left(\alpha_s^3\right)\right)\,,
\end{equation}
with $S_\epsilon=(4\pi)^{\eps}e^{-\eps\gamma_E}$, where $\gamma_E=-\Gamma'(1)$ is the Euler-Mascheroni constant. $\mu_0^2$ is the scale introduced in dimensional regularization to keep the coupling dimensionless in the QCD Lagrangian, and $\mu^2$ is the renormalization scale. In the following we set $\mu^2=\mu_0^2=1$.
For purely gluonic amplitudes, the coefficients of the QCD $\beta$ function are
\begin{equation}
    \beta_0=\frac{11N_C}{3}\,,\qquad
  \beta_1=\frac{17N_C^2}{3}\,.
\end{equation}
The renormalized amplitude is written as
\begin{equation}\label{eq:renormAmp}
  \mathcal{A}_R\big\vert_{\textrm{leading color}}=S_\epsilon^{-\frac{3}{2}}g_s^3\left(\mathcal{A}_R^{(0)}
  +\frac{\alpha_sN_C}{4\pi}\mathcal{A}_R^{(1)}
  +\left(\frac{\alpha_sN_C}{4\pi}\right)^2\mathcal{A}_R^{(2)}
  +\mathcal{O}(\alpha_s^3)
  \right)\,, 
\end{equation}
with the renormalized $\mathcal{A}_R^{(i)}$ related to the bare $\mathcal{A}^{(i)}$ as follows:
\begin{align}\begin{split}\label{eq:twoLoopUnRenorm}
    &\mathcal{A}_R^{(0)}=\mathcal{A}^{(0)}\,,\qquad
  \mathcal{A}_R^{(1)}=S_{\epsilon}^{-1}\mathcal{A}^{(1)}
  -\frac{3}{2}\frac{\beta_0}{N_C\epsilon}\mathcal{A}^{(0)}\,,\\
  &\mathcal{A}_R^{(2)}=
  S_{\epsilon}^{-2}\mathcal{A}^{(2)}
  -\frac{5}{2}\frac{\beta_0}{N_C\epsilon}S_{\epsilon}^{-1}
  \mathcal{A}^{(1)}
  +\left(\frac{15}{8}\left(\frac{\beta_0}{N_C\epsilon}\right)^2
  -\frac{3}{2}\frac{\beta_1}{N_C^2\epsilon}\right)\mathcal{A}^{(0)}\,.
\end{split}\end{align}

\subsection{Infrared Behavior}

Renormalization removes all poles of ultraviolet origin. The remaining poles in the renormalized amplitude are of infrared origin and can be predicted from the previous orders in the perturbative expansion \cite{Catani:1998bh,Sterman:2002qn,Becher:2009cu,Gardi:2009qi}:
\begin{align}\begin{split}\label{eq:catani}
    \mathcal{A}_R^{(1)}&={\bf I}^{(1)}(\epsilon)\mathcal{A}_R^{(0)}+\mathcal{O}(\epsilon^0)\,,\\
    \mathcal{A}_R^{(2)}&={\bf I}^{(2)}(\epsilon)\mathcal{A}_R^{(0)}+{\bf I}^{(1)}(\epsilon)
    \mathcal{A}_R^{(1)}+\mathcal{O}(\epsilon^0)\,.
\end{split}\end{align}
In the leading-color approximation, the color structure of loop corrections is the same as that of the leading-order contribution up to a factor of $N_C$ that was included in the perturbative expansion parameter, see eq.~\eqref{eq:renormAmp}. For a $n$-gluon amplitude, the operator~${\bf I}^{(1)}$ is then
\begin{equation}
  {\bf I}^{(1)}(\eps)=-\frac{e^{\gamma_E\eps}}{\Gamma(1-\epsilon)}
  \left(\frac{1}{\epsilon^2}+\frac{\beta_0}{2N_C\epsilon}\right)
  \sum_{i=1}^n\left(-s_{i,i+1}\right)^{-\epsilon}\,,
\end{equation}
where $s_{i,i+1}=(p_i+p_{i+1})^2$ with the indices defined cyclically. The operator~${\bf I}^{(2)}$ is
\begin{align}\begin{split}
  {\bf I}^{(2)}(\eps)=&
  -\frac{1}{2}{\bf I}^{(1)}(\eps){\bf I}^{(1)}(\eps)
  -\frac{\beta_0}{N_C\epsilon}{\bf I}^{(1)}(\eps)
  +\frac{e^{-\gamma_E\epsilon}\Gamma(1-2\epsilon)}{\Gamma(1-\epsilon)}
  \left(\frac{\beta_0}{N_C\epsilon}+\frac{67}{9}-\frac{\pi^2}{3}\right){\bf I}^{(1)}(2\epsilon)\\
  &+n\frac{e^{\gamma_E\epsilon}}{\epsilon\Gamma(1-\epsilon)}
  \left(\frac{\zeta_3}{2}+\frac{5}{12}+\frac{11\pi^2}{144}\right)\,.
\end{split}\end{align}
The poles of the bare amplitudes can be recovered from those of the renormalized amplitude by using eq.~\eqref{eq:twoLoopUnRenorm}. For amplitudes that are finite at one-loop (such as the all-plus and single-minus helicity configurations) the pole structure of the unrenormalized amplitude is particularly simple because the $1/\epsilon$ renormalization term in eq.~\eqref{eq:twoLoopUnRenorm} matches the $1/\epsilon$ term in the ${\bf I}^{(1)}$ operator \cite{Gehrmann:2015bfy}. For example, in the all-plus five-gluon case we have
\begin{equation}
  \mathcal{A}^{(2)}(1^+,2^+,3^+,4^+,5^+)=-
  \frac{1}{\epsilon^2}
  \sum_{i=1}^5\left(-s_{i,i+1}\right)^{-\epsilon}
  \mathcal{A}_R^{(1)}(1^+,2^+,3^+,4^+,5^+)
  +\mathcal{O}(\epsilon^0)\,.
\end{equation}

\bibliography{main.bib}

\end{document}